\documentclass[twocolumn,showpacs,prl,floatfix]{revtex4}

\usepackage{graphicx}
\usepackage{bm}
\usepackage{psfrag}
\usepackage{amsmath}
\usepackage{amssymb}
\usepackage{latexsym}
\usepackage{exscale}

\newcommand{\vect}[1]{\bm{#1}}
\newcommand{\ten}[1]{\mbox{\textbf{{\textsf{#1}}}}}

\newcommand{\veczero}{\mbox{\textbf{\textit{0}}}}

\newcommand{\sprod}{\!\cdot\!}
\newcommand{\tprod}{}
\newcommand{\vprod}{\!\times\!}
\newcommand{\trace}{\operatorname{Tr}}

\newcommand{\dif}{\mathrm{d}}
\newcommand{\mi}{\mathrm{i}}

\begin{document}

\title{Macroscopic quantum electrodynamics and duality}

\author{Stefan Yoshi Buhmann}
\author{Stefan Scheel}
\affiliation{Quantum Optics and Laser Science, Blackett Laboratory,
Imperial College London, Prince Consort Road,
London SW7 2AZ, United Kingdom}

\date{\today}

\begin{abstract}
We discuss under what conditions the duality between electric and
magnetic fields is a valid symmetry of macroscopic quantum
electrodynamics. It is shown that Maxwell's equations in the absence
of free charges satisfy duality invariance on an operator level,
whereas this is not true for Lorentz forces and atom--field couplings
in general. We prove that derived quantities like Casimir forces,
local-field corrected decay rates as well as van-der-Waals potentials
are invariant with respect to a global exchange of electric and
magnetic quantities. This exact symmetry can be used to deduce the
physics of new configurations on the basis of already established
ones.  
\end{abstract}

\maketitle


In the past, studies of quantum electrodynamic (QED) phenomena have
often been restricted to purely electric systems, because effects
associated with magnetic properties are considerably smaller for
materials occurring in nature. Two developments have recently
triggered an increased interest in such magnetic effects: The first
was the suggestion \cite{0477} and subsequent fabrication
\cite{0479} of artificial metamaterials with controllable electric
permittivity $\varepsilon$ and magnetic permeability $\mu$, where
left-handed materials (LHMs) with negative real parts of $\varepsilon$
and $\mu$ are of particular interest. As had already pointed out in
1968 \cite{0476}, the basis vectors of an electromagnetic wave
propagating inside such a medium form a left-handed triad, implying
negative refraction. Motivated by the progress in metamaterial
fabrication, researchers have intensively studied their potentials,
leading to proposals of a perfect lens with sub-wavelength resolution
\cite{0478} as well as cloaking devices \cite{0835} and predictions of
an unusual behaviour of the decay of one or two atoms in the presence
of LHMs \cite{0002,0737}.

Another, closely related motivation for considering magnetic systems
was due to the fact that dispersion forces \cite{0696} have gained an
increasing influence on micro-mechanical devices where they often lead
to undesired effects such as stiction \cite{0578}. The question
naturally arose whether LHMs could be exploited to modify or even
change the sign of dispersion forces. Forces on excited systems might
indeed be influenced by LHMs \cite{0828}. Ground-state forces are not
as easily manipulated because they depend on the medium response at
all frequencies, whereas the Kramers-Kronig relations imply that LHMs
can only be realized in limited frequency windows. However, the
controllable magnetic properties available in metamaterials can still
have a large impact on dispersion forces: The dispersion forces
between electric and magnetic atoms \cite{0089} or bodies \cite{0122}
differ both in sign and power laws from those between only electric
ones. On the search for repulsive dispersion forces, interactions of
electric/magnetic atoms \cite{0121}, plates \cite{0123,0134} and atoms
with plates \cite{0017,0012} have been studied; more complex problems
such as atom-atom interactions in the presence of a magneto-electric
bulk medium \cite{0829}, plate \cite{0009} or sphere \cite{0830} have
also been addressed. Reductions or even sign changes of the forces
have been predicted for such scenarios and have been attributed
primarily to large permeabilities rather than left-handed properties.

Metamaterials have thus considerably increased the parameter space at
one's disposal for manipulating QED phenomena. An efficient use of
this new freedom requires the formulation of general statements of
what might be achieved in principle. Working in this direction, upper
bounds for the strength of attractive and repulsive Casimir forces
have been formulated \cite{0134} and it has been proven that the force
between two mirror-symmetric purely electric bodies is always
attractive \cite{0503}. In the present Letter, we establish another
such general principle on the basis of the duality of Maxwell's
equations under an exchange of electric and magnetic fields
\cite{0844,0845}, also known as electric/magnetic reciprocity within a
generalised framework of classical electrodynamics \cite{0850}. In
particle physics, duality has been discussed as a symmetry of the
$\mathcal{N}=4$ supersymmetric Yang-Mills theory \cite{0860}. We will
prove its validity in the context of macroscopic QED \cite{0002,0696}
and show that under certain conditions, quantities such as decay rates
and dispersion forces are invariant with respect to a global exchange
of electric and magnetic properties. The parameter space to be
considered in the search for optimal geometries and materials will
thus be effectively halved.

We begin by verifying duality for macroscopic QED in the absence of
free charges and currents. We group the fields into dual pairs
$(\sqrt{\varepsilon_0}\hat{\vect{E}},\sqrt{\mu_0}\hat{\vect{H}})$,
$(\sqrt{\mu_0}\hat{\vect{D}},\sqrt{\varepsilon_0}\hat{\vect{B}})$ and
$(\sqrt{\mu_0}\hat{\vect{P}},
\sqrt{\varepsilon_0}\mu_0\hat{\vect{M}})$, so that Maxwell's
equations read
\begin{gather}
\label{d1}
\vect{\nabla}\sprod
 \begin{pmatrix}\sqrt{\mu_0}\hat{\vect{D}}\\
 \sqrt{\varepsilon_0}\hat{\vect{B}}\end{pmatrix}
=\begin{pmatrix}0\\ 0\end{pmatrix},\\
\label{d2}
\vect{\nabla}\vprod
 \begin{pmatrix}\sqrt{\varepsilon_0}\hat{\vect{E}}\\
 \sqrt{\mu_0}\hat{\vect{H}}\end{pmatrix}
 +\frac{\partial}{\partial t}
 \begin{pmatrix}0&1\\-1&0\end{pmatrix}
 \begin{pmatrix}\sqrt{\mu_0}\hat{\vect{D}}\\
 \sqrt{\varepsilon_0}\hat{\vect{B}}\end{pmatrix}
 =\begin{pmatrix}\veczero\\ \veczero\end{pmatrix}
\end{gather}
with
\begin{equation}
\label{d3}
 \begin{pmatrix}\sqrt{\mu_0}\hat{\vect{D}}\\
 \sqrt{\varepsilon_0}\hat{\vect{B}}\end{pmatrix}
 =\frac{1}{c}
 \begin{pmatrix}\sqrt{\varepsilon_0}\hat{\vect{E}}\\
 \sqrt{\mu_0}\hat{\vect{H}}\end{pmatrix}
 +\begin{pmatrix}\sqrt{\mu_0}\hat{\vect{P}}\\
 \sqrt{\varepsilon_0}\mu_0\hat{\vect{M}}\end{pmatrix}.
\end{equation}
Maxwell's equations are invariant under the general 
$\operatorname{SO}(2)$ duality transformation
\begin{equation}
\label{d4}
\begin{pmatrix}\vect{x}\\ \vect{y}\end{pmatrix}^\star
 =\mathcal{D}(\theta)\begin{pmatrix}\vect{x}\\ \vect{y}\end{pmatrix},
 \quad\mathcal{D}(\theta)
 =\begin{pmatrix}\cos\theta&\sin\theta\\ 
 -\sin\theta&\cos\theta\end{pmatrix},
\end{equation}
which may equivalently be expressed as a $\operatorname{U}(1)$
transformation when introducing complex Riemann--Silberstein fields
\cite{0844}. The invariance of Maxwell's equations under this
rotation can be verified by multiplying Eqs.~(\ref{d1})--(\ref{d3}) by
$\mathcal{D}(\theta)$ and using the fact that $\mathcal{D}(\theta)$
commutes with the symplectic matrix in Eq.~(\ref{d2}). Note that the
grouping into dual pairs is solely due to the mathematical structure
of the equations and is in contrast to the fact that $\hat{\vect{E}}$,
$\hat{\vect{B}}$ and $\hat{\vect{D}}$, $\hat{\vect{H}}$ are the pairs
of physically corresponding quantities.

For it to be a valid symmetry of the electromagnetic field, duality
must also be consistent with the constitutive relations. In the
presence of linear, local, isotropic, dispersing and absorbing media,
the constitutive relations in frequency space can be given as
\begin{multline}
\label{d5}
 \begin{pmatrix}\sqrt{\mu_0}\hat{\underline{\vect{D}}}\\
 \sqrt{\varepsilon_0}\hat{\underline{\vect{B}}}\end{pmatrix}
 =\frac{1}{c}\begin{pmatrix}\varepsilon&0\\
 0&\mu\end{pmatrix}
 \begin{pmatrix}\sqrt{\varepsilon_0}\hat{\underline{\vect{E}}}\\
 \sqrt{\mu_0}\hat{\underline{\vect{H}}}\end{pmatrix}\\
 +\begin{pmatrix}1&0\\0&\mu\end{pmatrix}
 \begin{pmatrix}\sqrt{\mu_0}\hat{\underline{\vect{P}}}_\mathrm{N}\\
 \sqrt{\varepsilon_0}\mu_0\hat{\underline{\vect{M}}}_\mathrm{N}
 \end{pmatrix},
\end{multline}
where $\varepsilon$ $\!=$ $\!\varepsilon(\vect{r},\omega)$ and
$\mu$ $\!=$ $\!\mu(\vect{r},\omega)$ denote the relative
electric permittivity and magnetic permeability of the media 
and $\hat{\vect{P}}_\mathrm{N}$ and $\hat{\vect{M}}_\mathrm{N}$ are
the noise polarisation and magnetisation which necessarily arise in
the presence of absorption. Invariance of the constitutive
relations (\ref{d5}) under the duality transformation requires that 
\begin{multline}
\label{d7}
\begin{pmatrix}\varepsilon^\star&0\\0&\mu^\star\end{pmatrix}
=\mathcal{D}(\theta)
\begin{pmatrix}\varepsilon&0\\0&\mu\end{pmatrix}
\mathcal{D}^{-1}(\theta)\\ 
=\begin{pmatrix}\varepsilon\cos^2\theta+\mu\sin^2\theta
 &(\mu-\varepsilon)\sin\theta\cos\theta\\
 (\mu-\varepsilon)\sin\theta\cos\theta
 &\varepsilon\sin^2\theta+\mu\cos^2\theta\end{pmatrix}.
\end{multline}

This condition is trivially fulfilled if \mbox{$\varepsilon$ $\!=$
$\!\mu$} (including both free space and the perfect lens,
\mbox{$\varepsilon$ $\!=$ $\!\mu$ $\!=$ $\!-1$} \cite{0478}), where  
duality is a continuous symmetry. For media with a non-trivial
impedance, the condition~(\ref{d7}) only holds for $\theta$ $\!=$
$\!n\pi/2$ with $n\in\mathbb{Z}$. The presence of such media thus 
reduces the continuous symmetry to a discrete symmetry with four
distinct members, whose group structure is that of $\mathbb{Z}_4$. For
$\theta$ $\!=$ $\!n\pi/2$, Eqs.~(\ref{d5}) and (\ref{d7}) imply the
transformations
\begin{gather}
\label{d10}
 \begin{pmatrix}\varepsilon\\ \mu\end{pmatrix}^\star
 =\begin{pmatrix}\cos^2\theta&\sin^2\theta\\
 \sin^2\theta&\cos^2\theta\end{pmatrix}
 \begin{pmatrix}\varepsilon\\ \mu\end{pmatrix},\\
\label{d11}
\begin{pmatrix}\sqrt{\mu_0}\hat{\underline{\vect{P}}}_\mathrm{N}\\
 \sqrt{\varepsilon_0}\mu_0\hat{\underline{\vect{M}}}_\mathrm{N}
 \end{pmatrix}^\star
=\begin{pmatrix}\cos\theta&\mu\sin\theta\\ 
 -\varepsilon^{-1}\sin\theta&
 \cos\theta\end{pmatrix}
\begin{pmatrix}\sqrt{\mu_0}\hat{\underline{\vect{P}}}_\mathrm{N}\\
 \sqrt{\varepsilon_0}\mu_0\hat{\underline{\vect{M}}}_\mathrm{N}
 \end{pmatrix}.
\end{gather}
 
Maxwell's equations~(\ref{d1}) and (\ref{d2}), together with the
constitutive relations~(\ref{d5}) for the electromagnetic field in the
absence of free charges and currents, are thus invariant under the
discrete duality transformations $\theta$ $\!=$ $\!n\pi/2$,
$n\in\mathbb{Z}$ given by Eqs.~(\ref{d4}), (\ref{d10}) and
(\ref{d11}). This is not only true for the equations of motion, but
clearly must also hold on a Hamiltonian level. To see this explicitly,
recall that the Hamiltonian of the medium-assisted field is given by 
\mbox{$\hat{H}_\mathrm{F}$ $\!=$ $\!\sum_{\lambda=e,m}\int\dif^3r
\int_0^\infty \dif\omega\,\hbar\omega\,
\hat{\vect{f}}_\lambda^\dagger(\vect{r},\omega)
\sprod\hat{\vect{f}}_\lambda(\vect{r},\omega)$} \cite{0002}
where the fundamental bosonic fields $\hat{\vect{f}}_\lambda$ are
related to the noise terms via
\begin{equation}
\label{d13}
\begin{pmatrix}\sqrt{\mu_0}\hat{\underline{\vect{P}}}_\mathrm{N}\\
 \sqrt{\varepsilon_0}\mu_0\hat{\underline{\vect{M}}}_\mathrm{N}
 \end{pmatrix}
=\sqrt{\frac{\hbar}{\pi c^2}}
\begin{pmatrix}\mi\sqrt{\operatorname{Im}\varepsilon}&0\\ 
 0&\sqrt{\operatorname{Im}\mu}/|\mu|\end{pmatrix}
\begin{pmatrix}\hat{\vect{f}}_e\\
 \hat{\vect{f}}_m\end{pmatrix}.
\end{equation}
Combining Eqs.~(\ref{d10}), (\ref{d11}) and (\ref{d13}), one finds
that the fundamental fields transform as
\begin{equation}
\label{d14}
\begin{pmatrix}\hat{\vect{f}}_e\\ \hat{\vect{f}}_m\end{pmatrix}^\star
 =\begin{pmatrix}\cos\theta
 &-\mi(\mu/|\mu|)\sin\theta\\
 -\mi(|\varepsilon|/\varepsilon)\sin\theta 
 &\cos\theta\end{pmatrix}
 \begin{pmatrix}\hat{\vect{f}}_e\\ \hat{\vect{f}}_m\end{pmatrix}
\end{equation}
for $\theta$ $\!=$ $\!n\pi/2$, so that $\hat{H}_\mathrm{F}^\star$
$\!=$ $\!\hat{H}_\mathrm{F}$. It is sufficient to focus on the single
duality transformation $\theta$ $\!=$ $\!\pi/2$ as summarised in
Tab.~\ref{Tab1}, which is a generator of the whole group.
\begin{table}[t]
\begin{tabular}{cccc}
\hline
Partners&\multicolumn{3}{c}{Transformation}\\
\hline
$\hat{\vect{E}}$, $\hat{\vect{H}}$:
&$\hat{\vect{E}}^\star=c\mu_0\hat{\vect{H}}$,&\hspace{1ex}&
$\hat{\vect{H}}^\star=-\hat{\vect{E}}/(c\mu_0)$\\
$\hat{\vect{D}}$, $\hat{\vect{B}}$:
&$\hat{\vect{D}}^\star=c\varepsilon_0\hat{\vect{B}}$,&\hspace{1ex}&
$\hat{\vect{B}}^\star=-\hat{\vect{D}}/(c\varepsilon_0)$\\
$\hat{\vect{P}}$, $\hat{\vect{M}}$:
&$\hat{\vect{P}}^\star=\hat{\vect{M}}/c$,&\hspace{1ex}&
$\hat{\vect{M}}^\star=-c\hat{\vect{P}}$\\
$\hat{\vect{P}}_A$, $\hat{\vect{M}}_A$:
&$\hat{\vect{P}}_A^\star=\hat{\vect{M}}_A/c$,&\hspace{1ex}&
$\hat{\vect{M}}_A^\star=-c\hat{\vect{P}}_A$\\
$\hat{\vect{d}}$, $\hat{\vect{m}}$:
&$\hat{\vect{d}}^\star=\hat{\vect{m}}/c$,&\hspace{1ex}&
$\hat{\vect{m}}^\star=-c\hat{\vect{d}}$\\
$\hat{\vect{P}}_\mathrm{N}$, $\hat{\vect{M}}_\mathrm{N}$:
&$\hat{\vect{P}}_\mathrm{N}^\star
=\mu\hat{\vect{M}}_\mathrm{N}/c$,&\hspace{1ex}
&$\hat{\vect{M}}_\mathrm{N}^\star
=-c\hat{\vect{P}}_\mathrm{N}/\varepsilon$\\
$\hat{\vect{f}}_e$, $\hat{\vect{f}}_m$:
&$\hat{\vect{f}}_e^\star
=-\mi(\mu/|\mu|)\hat{\vect{f}}_m$,&\hspace{1ex}
&$\hat{\vect{f}}_m^\star
=-\mi(|\varepsilon|/\varepsilon)\hat{\vect{f}}_e$\\
$\varepsilon$, $\mu$:
&$\varepsilon^\star=\mu$,&\hspace{1ex}
&$\mu^\star=\varepsilon$\\
$\alpha$, $\beta$:
&$\alpha^\star=\beta/c^2$,&\hspace{1ex}
&$\beta^\star=c^2\alpha$
\end{tabular}
\caption{%
\label{Tab1}
Effect of the duality transformation with $\theta$ $\!=$ $\!\pi/2$.}
\vspace{-3ex}
\end{table}%

Let us next turn our attention to Lorentz forces and the coupling of
the medium-assisted field to charged particles: We recall that the
operator Lorentz force on a neutral body occupying a volume $V$ can be
given as \cite{0696}
\begin{multline}
\label{d15x}
\hat{\vect{F}}
=\int_{\partial V}\dif\vect{A}\sprod\biggl\{
 \varepsilon_0\hat{\vect{E}}(\vect{r})
 \tprod\hat{\vect{E}}(\vect{r})
 +\frac{1}{\mu_0}\hat{\vect{B}}(\vect{r})
 \tprod\hat{\vect{B}}(\vect{r})\\
 -\frac{1}{2}\biggl[\varepsilon_0
 \hat{\vect{E}}^2(\vect{r})
 +\frac{1}{\mu_0}\hat{\vect{B}}^2(\vect{r})\biggr]\ten{I}
 \biggr\}\\
 -\varepsilon_0\,\frac{\dif}{\dif t}
 \int_V \dif^3r\,\hat{\vect{E}}(\vect{r})
 \vprod\hat{\vect{B}}(\vect{r})
\end{multline}
($\ten{I}$: unit tensor), while that on a neutral atom with
polarisation $\hat{\vect{P}}_A$ and magnetisation $\hat{\vect{M}}_A$
reads \cite{0696,0008}
\begin{multline}
\label{d16x}
\hat{\vect{F}}
=\vect{\nabla}_{\!\!{A}}\int\dif^3r\,\Bigl[
 \hat{\vect{P}}_A(\vect{r})\sprod\hat{\vect{E}}(\vect{r})
 +\hat{\vect{M}}_A(\vect{r})\sprod\hat{\vect{B}}(\vect{r})\\
+\hat{\vect{P}}_A(\vect{r})\vprod\dot{\hat{\vect{r}}}_{A}
 \sprod\hat{\vect{B}}(\vect{r})\Bigr]
 +\frac{\dif}{\dif t}\int\dif^3r\,
 \hat{\vect{P}}_A(\vect{r})\vprod\hat{\vect{B}}(\vect{r}).
\end{multline}
The coupling of one or more atoms to the medium-assisted
electromagnetic field can in the multipolar coupling scheme be
implemented via \cite{0696,0009}
\begin{multline}
\label{d17x}
\hat{H}_{A\mathrm{F}}
=-\int\dif^3r\,\Bigl[
 \hat{\vect{P}}_A(\vect{r})\sprod\hat{\vect{E}}(\vect{r})
 +\hat{\vect{M}}_A(\vect{r})\sprod\hat{\vect{B}}(\vect{r})\\
 +m_A^{-1}\hat{\vect{P}}_A(\vect{r})\vprod\hat{\vect{p}}_{A}
 \sprod\hat{\vect{B}}(\vect{r})\Bigr],
\end{multline}
when neglecting diamagnetic interactions. Using the transformation
behaviour given in Tab.~\ref{Tab1}, it is immediately clear that
neither the Lorentz forces on bodies or atoms nor the atom-field
interactions are duality invariant on an operator level. Even for
atoms and bodies at rest with time-independent fields, duality
invariance is prohibited by the unavoidable noise polarisation and
magnetisation in the constitutive relations~(\ref{d5}).

That said, we will show that effective quantities derived from the
above operator Lorentz forces and atom--field couplings do obey
duality invariance when considering atoms and bodies at rest and not
embedded in a medium. In particular, we will consider the Casimir
force \cite{0198}
\begin{multline}
\label{d15y}
\vect{F}=-\frac{\hbar}{\pi}\int_{0}^{\infty} \dif\xi
 \int_{\partial V}\dif\vect{A}\sprod\Bigl\{
 \ten{G}_{ee}^{(1)}(\vect{r},\vect{r},\mi\xi)
 +\ten{G}_{mm}^{(1)}(\vect{r},\vect{r},\mi\xi)\\
 -{\textstyle\frac{1}{2}}\trace\Bigl[
 \ten{G}_{ee}^{(1)}(\vect{r},\vect{r},\mi\xi)
 +\ten{G}_{mm}^{(1)}(\vect{r},\vect{r},\mi\xi)
 \Bigr]\ten{I}\Bigr\},
\end{multline}
the single- and two-atom vdW potentials \cite{0696,0012,0831}
\begin{multline}
\label{d17y}
U(\vect{r}_{\!A})
 =\frac{\hbar}{2\pi\varepsilon_0}
 \int_0^\infty\dif\xi\,\Bigl[\alpha(\mi\xi)
 \trace\ten{G}_{ee}^{(1)}(\vect{r}_{\!A},\vect{r}_{\!A},\mi\xi)\\
 +\frac{\beta(\mi\xi)}{c^2}\,\trace
 \ten{G}_{mm}^{(1)}(\vect{r}_{\!A},\vect{r}_{\!A},\mi\xi)\Bigr]
\end{multline}
and
\begin{align}
\label{d17z}
&U(\vect{r}_{\!A},\vect{r}_{\!B})
 =-\frac{\hbar}{2\pi\varepsilon_0^2}\int_0^\infty\dif\xi
 \nonumber\\
&\times\trace\Bigl\{\alpha_A(\mi\xi)\alpha_B(\mi\xi) 
 \ten{G}_{ee}(\vect{r}_{\!A},\vect{r}_{\!B},\mi\xi)
 \sprod\ten{G}_{ee}(\vect{r}_{\!B},\vect{r}_{\!A},\mi\xi)
 \nonumber\\
&+\alpha_A(\mi\xi)\,\frac{\beta_B(\mi\xi)}{c^2}\, 
 \ten{G}_{em}(\vect{r}_{\!A},\vect{r}_{\!B},\mi\xi)
 \sprod\ten{G}_{me}(\vect{r}_{\!B},\vect{r}_{\!A},\mi\xi)
 \nonumber\\
&+\frac{\beta_A(\mi\xi)}{c^2}\,\alpha_B(\mi\xi)
 \ten{G}_{me}(\vect{r}_{\!A},\vect{r}_{\!B},\mi\xi)
 \sprod\ten{G}_{em}(\vect{r}_{\!B},\vect{r}_{\!A},\mi\xi)
 \nonumber\\
&+\frac{\beta_A(\mi\xi)}{c^2}\,\frac{\beta_B(\mi\xi)}{c^2}\,
 \ten{G}_{mm}(\vect{r}_{\!A},\vect{r}_{\!B},\mi\xi)
 \sprod\ten{G}_{mm}(\vect{r}_{\!B},\vect{r}_{\!A},\mi\xi)\Bigr\}
\end{align}
($\alpha$, $\beta$: atomic polarisability, magnetisability)
and the atomic decay rate \cite{0002,0832}
\begin{multline}
\label{d17q}
\Gamma_n(\vect{r}_{A})=\frac{2}{\hbar\varepsilon_0}\,\sum_{k<n}
 \biggl[\vect{d}_{kn}\sprod
 \operatorname{Im}\,\ten{G}_{ee}(\vect{r}_{\!A},\vect{r}_{\!A},
 \omega_{nk})\sprod\vect{d}_{nk}\\
 +\frac{\vect{m}_{kn}}{c}\,\sprod
 \operatorname{Im}\,\ten{G}_{mm}(\vect{r}_{\!A},\vect{r}_{\!A},
 \omega_{nk})\sprod\frac{\vect{m}_{nk}}{c}\biggr]
\end{multline}
($|n\rangle$: initial atomic state, $\omega_{nk}$: atomic transition
frequencies; $\vect{d}_{nk}$, $\vect{m}_{nk}$: electric, magnetic
dipole matrix elements).  Here, $\ten{G}^{(1)}$ is the scattering part
of the classical Green tensor, where a left index $e$, $m$ indicates
that $\ten{G}$ is multiplied by $\mi\omega/c$ $\!=$ $\!-\xi/c$ or
$\vect{\nabla}\vprod$ from the left and a right index $e$, $m$ denotes
multiplication with $\mi\omega/c$ $\!=$ $\!-\xi/c$ or
$\vprod\overleftarrow{\vect{\nabla}}'$ from the right. The Casimir
force and the single-atom vdW force are the ground-state averages of
the above operator Lorentz forces, while the atomic potentials and
rates follow from the atom--field coupling. 

To prove the duality invariance of the above quantities
(\ref{d15y})--(\ref{d17q}), we note that the Casimir force depends
solely on the classical Green tensor
\begin{equation}
\label{d18x}
\biggl[\vect{\nabla}\vprod
 \,\frac{1}{\mu(\vect{r},\omega)}\,\vect{\nabla}\vprod
 \,-\,\frac{\omega^2}{c^2}\,\varepsilon(\vect{r},\omega)\biggr]
 \ten{G}(\vect{r},\vect{r}',\omega)
 =\bm{\delta}(\vect{r}-\vect{r}'),
\end{equation}
while vdW forces and decay rates also depend on $\alpha$, $\beta$,
$\hat{\vect{d}}$ and $\hat{\vect{m}}$. While the transformation
behaviour of the latter quantities under duality follows immediately
from that of $\varepsilon$, $\mu$, $\hat{\vect{P}}_A$ and
$\hat{\vect{M}}_A$ (see Tab.~\ref{Tab1}), the transformed Green
tensor, which is the solution to Eq.~(\ref{d18x}) with $\varepsilon$
and $\mu$ exchanged, can be determined as follows: We first note that
Maxwell's equations~(\ref{d1}), (\ref{d2}) together with the
constitutive relations~(\ref{d5}) are uniquely solved by \cite{0002}
\begin{align}
\label{d15}
&\hat{\underline{\vect{E}}}(\vect{r},\omega)
=-\varepsilon_0^{-1}\int\dif^3r'\,
 \ten{G}_{ee}(\vect{r},\vect{r}',\omega)
 \sprod\hat{\underline{\vect{P}}}_\mathrm{N}(\vect{r}',\omega)
 \nonumber\\
&\quad-c\mu_0\int\dif^3r'\,\ten{G}_{em}(\vect{r},\vect{r}',\omega)
 \sprod\hat{\underline{\vect{M}}}_\mathrm{N}(\vect{r}',\omega),
\end{align}
\begin{align} 
\label{d16}
&\hat{\underline{\vect{B}}}(\vect{r},\omega)
=-c\mu_0\int\dif^3r'\,
 \ten{G}_{me}(\vect{r},\vect{r}',\omega)
 \sprod\hat{\underline{\vect{P}}}_\mathrm{N}(\vect{r}',\omega)
 \nonumber\\
&\quad-\mu_0\int\dif^3r'\,
 \ten{G}_{mm}(\vect{r},\vect{r}',\omega)
 \sprod\hat{\underline{\vect{M}}}_\mathrm{N}(\vect{r}',\omega),
\end{align} 
\begin{align}
\label{d17}
&\hat{\underline{\vect{D}}}(\vect{r},\omega)
 =-\frac{\varepsilon(\vect{r},\omega)}{c}\,
 \int\dif^3r'\,\ten{G}_{em}(\vect{r},\vect{r}',\omega)
 \sprod\hat{\underline{\vect{M}}}_\mathrm{N}(\vect{r}',\omega)
 \nonumber\\
&-\int\!\dif^3r'\!\biggl[\varepsilon(\vect{r},\omega)
\ten{G}_{ee}(\vect{r},\vect{r}',\omega)
 -\bm{\delta}(\vect{r}\!-\!\vect{r}')\biggr]
 \sprod\hat{\underline{\vect{P}}}_\mathrm{N}(\vect{r}',\omega),
\end{align}
\begin{align}
\label{d18}
&\hat{\underline{\vect{H}}}(\vect{r},\omega)
=-\frac{c}{\mu(\vect{r},\omega)}\int\dif^3r'\,
 \ten{G}_{me}(\vect{r},\vect{r}',\omega)
 \sprod\hat{\underline{\vect{P}}}_\mathrm{N}(\vect{r}',\omega)
 \nonumber\\
&-\int\dif^3r'\biggl[\frac{\ten{G}_{mm}(\vect{r},\vect{r}',\omega)}
 {\mu(\vect{r},\omega)}
 +\bm{\delta}(\vect{r}\!-\!\vect{r}')\biggr]
 \sprod\hat{\underline{\vect{M}}}_\mathrm{N}(\vect{r}',\omega).
\end{align}
The invariance of Maxwell's equations implies that this solution
remains valid after applying the duality transformation. Taking
duality transforms of Eqs.~(\ref{d15}) and (\ref{d16}), the unknown
transformed Green tensor appears on the rhs of these equations,
whereas the transformations of all other quantities occurring in the
equations can be determined with the aid of Tab.~\ref{Tab1}. After
using Eqs.~(\ref{d15})--(\ref{d18}) to express the resulting fields on
the lhs in terms of $\hat{\underline{\vect{P}}}_\mathrm{N}$ and
$\hat{\underline{\vect{M}}}_\mathrm{N}$ and equating coefficients, one
obtains the following transformation rules:
\begin{align}
\ten{G}_{ee}^\star(\vect{r},\vect{r}',\omega)
 =&\;\mu^{-1}(\vect{r},\omega)
 \ten{G}_{mm}(\vect{r},\vect{r}',\omega)
 \mu^{-1}(\vect{r}',\omega)
 \nonumber\\
\label{d19}
&+\mu^{-1}(\vect{r},\omega)
 \bm{\delta}(\vect{r}\!-\!\vect{r}'),\\
\label{d20}
\ten{G}_{em}^\star(\vect{r},\vect{r}',\omega)
=&-\mu^{-1}(\vect{r},\omega)
 \ten{G}_{me}(\vect{r},\vect{r}',\omega)
 \varepsilon(\vect{r}',\omega),\\
\label{d21}
\ten{G}_{me}^\star(\vect{r},\vect{r}',\omega)
=&-\varepsilon(\vect{r},\omega)
 \ten{G}_{em}(\vect{r},\vect{r}',\omega)
 \mu^{-1}(\vect{r}',\omega),\\
\ten{G}_{mm}^\star(\vect{r},\vect{r}',\omega)
=&\;\varepsilon(\vect{r},\omega)
 \ten{G}_{ee}(\vect{r},\vect{r}',\omega)
 \varepsilon(\vect{r}',\omega)\nonumber\\
&-\varepsilon(\vect{r},\omega)\bm{\delta}(\vect{r}\!-\!\vect{r}').
\label{d22}
\end{align}

The duality invariance of dispersion forces and decay rates follows
immediately. Using Eqs.~(\ref{d19}) and (\ref{d22}) and noting that
the $\delta$ function does not contribute to the scattering
part of the Green tensor, it is seen that the Casimir
force~(\ref{d15y}) on a body is unchanged when globally exchanging
$\varepsilon$ and $\mu$, provided that the body is located
in free space. The duality invariance of the vdW
potentials~(\ref{d17y}) and (\ref{d17z}) also follows from the
transformation rules~(\ref{d19})--(\ref{d22}). This invariance with
respect to a simultaneous exchange $\varepsilon\leftrightarrow\mu$ and
$\alpha\leftrightarrow\beta/c^2$ again only holds if
$\varepsilon(\vect{r}_{\!A/B})$ $\!=$ $\mu(\vect{r}_{\!A/B})$ $\!=$
$\!1$. In contrast to the Casimir force, this does not mean that the
atom has to be located in vacuum, but merely implies that for atoms
embedded in media, local-field corrections must be included via the
real-cavity model in order to insure invariance \cite{0739}. 

Duality invariance can be used to obtain the full functional
dependence of dispersion forces in given scenarios on the atomic and
medium parameters from knowledge of the respective dual scenario. For
instance, it has recently been shown that in the retarded limit the
vdW potential of two polarisable atoms reads
$U(r_{AB})=-1863\hbar c\alpha_A\alpha_B\varepsilon^2%
/[64\pi^3\varepsilon_0^2\sqrt{\varepsilon\mu}%
(2\varepsilon+1)^4r_{AB}^7]$ when including local-field
corrections \cite{0739}. Making the replacements
$\alpha\rightarrow\beta/c^2$, $\varepsilon\leftrightarrow\mu$, one can
immediately infer 
$U(r_{AB})=-1863\hbar c\mu_0^2\beta_A\beta_B\mu^2%
/[64\pi^3\sqrt{\varepsilon\mu}(2\mu+1)^4r_{AB}^7]$ for magnetisable
atoms. The utility of this principle becomes even more apparent for
complex problems like the interaction of two atoms in the presence
of a magneto-electric object \cite{0009,0830}. Finally, using the fact
that two purely electric, mirror-symmetric bodies always attract
\cite{0503}, we can immediately conclude that so do two purely
magnetic ones.

In addition, Eqs.~(\ref{d19}) and (\ref{d22}) imply the duality
invariance of the decay rate~(\ref{d17q}) since the $\delta$ functions
do not contribute to the imaginary part of the Green tensor; again,
local-field corrections have to be included for atoms embedded in
media. This symmetry can be exploited, e.g., to obtain magnetically
driven spin-flip rates of atoms in specific environments from known
electric-dipole driven decay rates. 

In conclusion, we have shown that dispersion forces on atoms and
bodies as well as decay rates of atoms are duality invariant, provided
that the bodies are located in free space at rest and that
local-field corrections are taken into account when considering
(stationary) atoms embedded in a medium. The established symmetry
operation of globally exchanging electric and magnetic body and atom
properties is a powerful tool for obtaining new results on the basis
of already established ones. The invariance can easily be extended to
other effective quantities of macroscopic QED such as frequency
shifts, heating rates or energy transfer rates. 


\acknowledgments
This work was supported by the Alexander von Humboldt Foundation and
the UK Engineering and Physical Sciences Research Council. S.Y.B. is
grateful to F.W.~Hehl and T.~K\"astner for discussions.


\end{document}